# Optimization of the growth of the van der Waals materials $Bi_2Se_3$ and $(Bi_{0.5}In_{0.5})_2Se_3$ by molecular beam epitaxy

*Zhengtianye Wang, Stephanie Law\**

Material Science and Engineering Department, 201 DuPont Hall, University of Delaware, Newark, Delaware, 19716, U.S.A.

Abstract: The naturally existing chalcogenide $Bi_2Se_3$ is topologically nontrivial due to the band inversion caused by strong spin-orbit coupling inside the bulk of the material. The surface states are spin polarized, protected by the time-inversion symmetry, and thus robust to the scattering caused by non-magnetic defects. A high purity topological insulator thin film can be easily grown via molecular beam epitaxy (MBE) on various substrates to enable novel electronics, optics, and spintronics applications. However, the unique surface state properties have historically been limited by the film quality, which is evaluated by crystallinity, surface morphology, and transport data. Here we propose and investigate different MBE growth strategies to improve the quality of $Bi_2Se_3$ thin films grown by MBE. Based on the surface passivation status, we have classified the substrates into two categories, self-passivated or unpassivated, and determine the optimal growth mechanisms on the representative sapphire and GaAs, respectively. For $Bi_2Se_3$ on GaAs, the surface passivation status determines the dominant growth mechanism. In the end, growths of topological trivial insulator (Bi$_{0.5}$In$_{0.5}$)$_2$Se$_3$ (BIS) on GaAs are investigated follow the protocols proposed.

1. **Introduction**

Characterizing electronic materials by the topology of their band structure has been a topic of significant interest in condensed matter and materials physics for the past decade. Materials can be categorized into different topological classes based on the $\mathbb{Z}_2$ invariance of the band structure. Topological insulators (TI) are among the materials with nonzero $\mathbb{Z}_2$ invariance[1]. At the boundary of a TI and a topologically trivial material, metallic helical edge states emerge due to the difference in band topology. These linearly-dispersed edge states often reside in the topological material bulk band gap, and electrons occupying these states exhibit spin-momentum locking: electrons moving in opposite directions have opposite spin polarizations. TIs are thus candidates for spintronic devices[2], quantum computation[3], and optoelectronics especially in the THz frequency band[4,5]. Bi$_2$Se$_3$, one of the most well-studied three-dimensional TIs, is often used for these applications because of its simple chemical form, relatively large bandgap, and ease of synthesis[6,7]. A perfect Bi$_2$Se$_3$ crystal would show metallic conducting surface states while the bulk remains insulating. To preserve and to take full advantage of the surface states, researchers have explored various ways to obtain crystalline thin films, ranging from top-down methods like mechanical exfoliation[8–11] to bottom-up methods like epitaxial growth via molecular beam epitaxy (MBE)[12–14], physical vapor deposition (PVD)[15], chemical vapor deposition (CVD)[16,17], etc.. MBE is often used for chalcogenide-based TI growth because it produces high purity films, enables precise control of the growth parameters including film thickness and doping level, and it can be used to create atomically-sharp interfaces in TI-based heterostructures[18–20]. MBE can also be used to synthesize pristine topological insulator nanoparticles[21,22]. Unfortunately, applications of the $Bi_2Se_3$ surface states have been limited by the poor film quality: intrinsic point defects (such as selenium vacancies or metal/chalcogenide antisite defects) may introduce unwanted doping, causing the Fermi level to be pinned in the conduction band, which will obscure the signals from the surface states[12]; and twin domains (also called antiphase domains) often form due to the 3-fold crystal symmetry, introducing dislocations in the film which serve as extra scattering paths for the surface states, thus limiting the surface state electron mobility[23]. A variety of MBE-based techniques have been used to reduce the doping and to improve the film crystallinity[12,24]. In this paper, we will do a head-to-head comparison of several growth strategies that have previously been tried on TIs, transition metal dichalcogenides, or III-V semiconductors. It should be noted that this is not meant to be an exhaustive comparison of every growth technique. Instead, we have chosen commonly-used techniques in a variety of categories. By growing the films in the same chamber on different substrates, we are able to fairly compare these growth methods as well as understand the differences when growing on inert versus unpassivated substrates. We evaluate the sample quality through x-ray studies of the crystal structure, scanning probe studies of the film surface morphology, and room-temperature transport measurements to determine the optimal growth procedures. We will show that due to the growth mechanism of van der Waals (vdW) epitaxy, $Bi_2Se_3$ with rhombohedral structure (trigonal unit cell) can be grown on GaAs (001) with cubic zincblende structure, although the early growth phase is strongly affected by the dangling bonds. Finally, we will discuss future directions for further improving the film quality.

The vdW materials of interest have layered atomic structures, so thin films of these materials grown by MBE follow the vdW epitaxy mechanism in which the vdW gaps are parallel to the substrate[25,26]. Unlike in the conventional III-V material systems where adatoms form covalent bonds with the substrate thus requiring strict lattice matching, in vdW epitaxy, the epitaxial material is believed to be linked to the substrate via vdW forces. The lattice-matching between the film and substrate is thus somewhat less critical. $Bi_2Se_3$, for example, has a layered structure at room temperature and ambient pressure, as shown in Figure 1(a). Within the layer, the selenium and bismuth atoms are covalently bonded in the order of 'Se-Bi-Se-Bi-Se', thus the entire layer is often referred as a quintuple layer (QL). These QLs are then connected to each other and to the substrate via vdW forces. However, when the lattice mismatch is nonnegligible, the $Bi_2Se_3$-substrate interface become defective and strained. For example, the in-plane lattice constant of sapphire (0001) is a=4.76 Å[27] while for $Bi_2Se_3$, it is a=4.14 Å[28,29]. The lattice mismatch is nearly 15%. This mismatch coupled with the vdW interaction means that the adatoms supplied to the substrate do not have a clear lowest-energy position to occupy, which eventually result in disorder or a distortion of the first few QLs. Fortunately, the overall strain relaxes quickly[30], but this does not mean that the substrate has no influence on the film quality. In fact, for some substrate materials, the pre-growth surface conditions play a large role in determining the resulting film quality. In the following, we will discuss in detail how the different types of substrates affect the morphology and electronic properties of $Bi_2Se_3$ crystalline films grown via MBE.

$Bi_2Se_3$ has been successfully grown on many substrates including silicon (Si)[31,32], III-Vs like gallium arsenide (GaAs)[33] and indium phosphide (InP)[23,34,35], sapphire (α-$Al_2O_3$)[12,36], hexagonal boron nitride (h-BN)[37], graphene[38], strontium titanium oxide ($SrTiO_3$)[39,40], and rare earth iron garnets like yttrium iron garnet (YIG)[41] and thulium iron garnet (TmIG)[42,43]. For the purposes of this paper, we divided the substrates into two categories: substrates with no dangling bonds like epi-ready sapphire, or substrates with dangling bonds and/or surface reconstructions, like GaAs.

For GaAs substrates used in a III-V system, the surface oxide layer is often thermally removed by heating the substrate to the oxide desorption temperature. An arsenic flux is supplied during the heating to compensate for the arsenic outgassing from the wafer and to prevent the formation of gallium droplets. In our chalcogenide MBE, we used selenium flux to replace the arsenic flux in the deoxidation process. Apart from the valence electron difference, selenium and arsenic are next to each other in the periodic table, so they have similar atomic masses and chemical reactivity. Selenium passivation of GaAs (001) has already been investigated at length in the growth of chalcogenide semiconductors like ZnSe[44–47]. However, different surface reconstructions, including 4×3, 2×3, and 2×1 have been observed in RHEED when the substrate is exposed to selenium flux at different substrate temperatures[48]. For selenium exposure at high substrate temperature (>200°C) the most reported reconstruction in both experiments and theoretical calculations is 2×1[49–52]. In this configuration, the first layer of arsenic is completely substituted by selenium, which is connected to two gallium atoms in the second layer; the third layer is mostly replaced (>90%) by selenium while the fifth layer partially replaced[50]. In these studies, the GaAs wafer is either deoxidized inside the MBE with an arsenic flux before transfer to a chalcogenide system, or a GaAs epitaxial layer was grown[49,52]. An arsenic-rich 2×4 surface is present before any treatment. However, for selenium exposure at room temperature, the surface arsenic atoms remain and only As-Se bonds are formed[49,53], after which a selenium overlayer follows as deposition continues. It is reported that the surface structure changed from 2×4 to 2×1 while lowering the substrate temperature, before applying selenium flux[49]. After the selenium deposition, the film can be further heat-treated to show a stable 4×1[54] or 2×3[55] structure. A similar 1×1 to 4×1 reconstruction has been seen in GaAs (001) substrate passivated with aqueous selenium-based reagent, when heated up to 580°C[56]. In regard to the deoxidation and passivation process that we use, we observe a constant 2×4 reconstruction prior to the Bi$_2$Se$_3$ growth, which has simplified our growth process and analysis of the dynamics of vdW epitaxy on GaAs. Combined with the aforementioned selenium passivation results and our method of deoxidation, it is reasonable to assume that the topmost layer is dominated by selenium. We cannot completely rule out selenium-arsenic bonding, but it is unlikely to happen at such high temperature. We anticipate that gallium selenides will form at the surface of the deoxidized GaAs at the temperature we use. It is safe to assume that the substrate has some unsatisfied dangling atomic bonds which may lead to misalignments (i.e. different orientations and bonding mechanism) of the first few QLs of Bi$_2$Se$_3$ grown on top. These misalignments may eventually serve as origins of the defects that limit the performance of the TI thin films, so understanding the substrate pre-treatment is crucial for the growth of high-quality films.

To mitigate the high density of defects at the Bi$_2$Se$_3$-substrate interface, a (Bi$_{0.5}$In$_{0.5}$)$_2$Se$_3$ (BIS) buffer layer is often added to preserve the bottom surface states and relax the strain from the substrate[24,57]. BIS is a topologically-trivial band insulator that has the same rhombohedral crystal lattice as Bi$_2$Se$_3$, but slightly shifted lattice constants (a$_{Bi2Se3}$=4.14Å, a$_{BIS}$~4.02Å)[58–60]. By introducing the BIS buffer layer between the film and the sapphire substrate, the carrier density of the Bi$_2$Se$_3$ layer is reduced by half compared to films directly grown on the bare substrate[24]. This indicates that when directly grown, at least half of the additional carriers arise from defects at the Bi$_2$Se$_3$ interface with the sapphire substrate. BIS, on the other hand, can act as a buffer to bury these interfacial defects. This buffer layer strategy can be used on other substrates as well. Finally, due to its topologically-trivial nature, BIS can also be used as a capping layer for Bi$_2$Se$_3$ to protect the top surface and as a separation layer for vertically stacked Bi$_2$Se$_3$ multilayer structures[4,61,62].

In the following sections, we will first introduce five different growth protocols, indexed A to E, for $Bi_2Se_3$ directly grown on top of representative substrates from each class: c-plane sapphire with no dangling bonds and (001) oriented undoped GaAs with selenium-passivated dangling bonds. We first categorize and number the samples based on the growth strategy. Then in each subsection differentiated by substrate, we show the characterization results of these films. The growth is characterized by *in-situ* reflection high energy electron diffraction (RHEED) to monitor the surface reconstruction and roughness. The films are characterized *ex-situ* with Hall effect measurements, atomic force microscopy (AFM), and x-ray diffraction (XRD). At last, in the discussion section, we will synthesize the results of the five growth strategies and compare our findings with previous reports. We close the discussion by elaborating on how the film quality is affected by the growth process, as well as the limitations of each strategy.

## 2. Method

All the $Bi_2Se_3$ or BIS films are grown using a Veeco GENxplor R&D MBE system. Knudson effusion cells are used for bismuth (UMC Corp., 6N purity grade), and indium (UMC Corp., 7N purity grade) sources. For the selenium cell (UMC Corp., 6N purity grade), the base temperature is kept between 280-290°C while an extra cracker zone is installed to heat the flux to 900°C to break the large selenium molecules into smaller, more reactive species for better incorporation into the film. The 10mm×10mm×0.5mm (0001) plane sapphire substrates (MTI Corporation) or a quarter of 2-inch (001) plane single side polished GaAs with 0.5mm thickness (Wafer Technology Ltd.) are first baked in the load lock chamber at 200°C for 10 hours and then transferred to growth chamber with base pressure below $1.0 \times 10^{-9}$ Torr. To further clean the sapphire substrate, the substrate is directly heated to 650°C with ramp rate of 20°C/min and held at this temperature for five minutes to degas. Then it is cooled back to the growth temperature (depending on the strategy) for $Bi_2Se_3$. In the case of GaAs, we first heated the substrate directly to 300°C with no selenium flux. We then started the selenium flux (same beam equivalent pressure (BEP) that is used for the $Bi_2Se_3$ growth where the BEP is measured by an ionization gauge with a thoria-coated iridium filament placed in the substrate position). We then heated the substrate to 760°C to thermally desorb the surface oxide layer. The ramp rate is kept at 20°C/min. Finally, we increased the GaAs substrate temperature to 770°C with the ramp rate of 5°C/min. We observe a 2×4 GaAs surface reconstruction at 770°C. After the deoxidation, the substrate is cooled to the growth temperature while the selenium flux is maintained. During $Bi_2Se_3$ growth, the selenium:bismuth (or selenium:(bismuth + indium) for BIS) ratio is kept around 80-100 as measured by the flux gauge. The selenium overpressure is kept unchanged for the GaAs deoxidation, $Bi_2Se_3$ film growth, annealing, and decomposition point testing. The thickness of all samples discussed in this paper are kept at 50 nm unless otherwise specified. The deposition rate for $Bi_2Se_3$ is kept at 0.74nm/min unless specified while the deposition rate for BIS is about 1nm/min. After growing the 50 nm film using any method, the sample is cooled down to 200°C with the ramp rate of 20°C/min under the selenium flux to prevent selenium outgassing from the sample surface. All substrate temperatures are measured by a non-contact thermocouple.

The room temperature Hall measurement is taken using the van der Pauw configuration within 20 minutes after taking the samples out of the MBE system. For each sample, we take three rounds of measurements and obtain an average and error bars. The samples are then vacuum packed for storage. Wide angle XRD scans are done with a Bruker D8 XRD, with 0.03° step and 0.3s per each step. Ψ scans of the $\{10\bar{1}5\}$ plane of $Bi_2Se_3$ are taken with a Rigaku Ultima IV XRD. AFM scans are taken with a Veeco Dimension 3100V scanning probe microscope with an antimony coated silicon tip (Bruker, NCHV-A) over a 5 $\mu$m×5 $\mu$m area near the center of the sample with 1Hz scan rate. At the end of Section 3, we present Table 1 and Table 2 to summarize all the samples under consideration, categorized by the growth method, and list the Hall measurement results for easy comparison.

3. Results

To begin, we will first discuss how we calibrate the non-contact thermocouple reading of the substrate heater using the thermal decomposition point of the $Bi_2Se_3$ film. This calibration is crucial to account for differences in thermocouple reading and physical temperature between MBE systems or between different substrate holders within the same system. This decomposition temperature serves as a good reference point for the temperature study of $Bi_2Se_3$ growth dynamics (and in principle, also for other chalcogenide materials that grow near that temperature). We first deposit 2QL of $Bi_2Se_3$ at 325°C and then heat the film to 425°C under a selenium overpressure that is the same as when we grow the film. We then heat the substrate by 5°C and hold at each temperature for 5 minutes to observe the RHEED pattern. We find that when the substrate reaches ~435°C, the RHEED pattern for $Bi_2Se_3$ starts to vanish and we observe more of the sapphire RHEED pattern. We define this thermocouple temperature as the $Bi_2Se_3$ decomposition point. All growth temperatures read by the thermocouple are then defined relative to the decomposition point. Since this is a physical transition, it is independent of MBE system configuration or sample holder. The disappearance of $Bi_2Se_3$ may be caused by the increase of the $Bi_2Se_3$ surface tension on sapphire, so that it will not wet the substrate anymore. It is shown in Ref.[63] that if we insert a 10nm BIS seed layer before growing $Bi_2Se_3$ on sapphire, the $Bi_2Se_3$ growth temperature can be raised up to 500°C. The GaAs deoxidation temperature can also be used as a physical reference for samples grown on GaAs, though it is not as useful as the thermal degradation point since it is hundreds of degrees higher than normal Bi$_2$Se$_3$ growth temperatures.

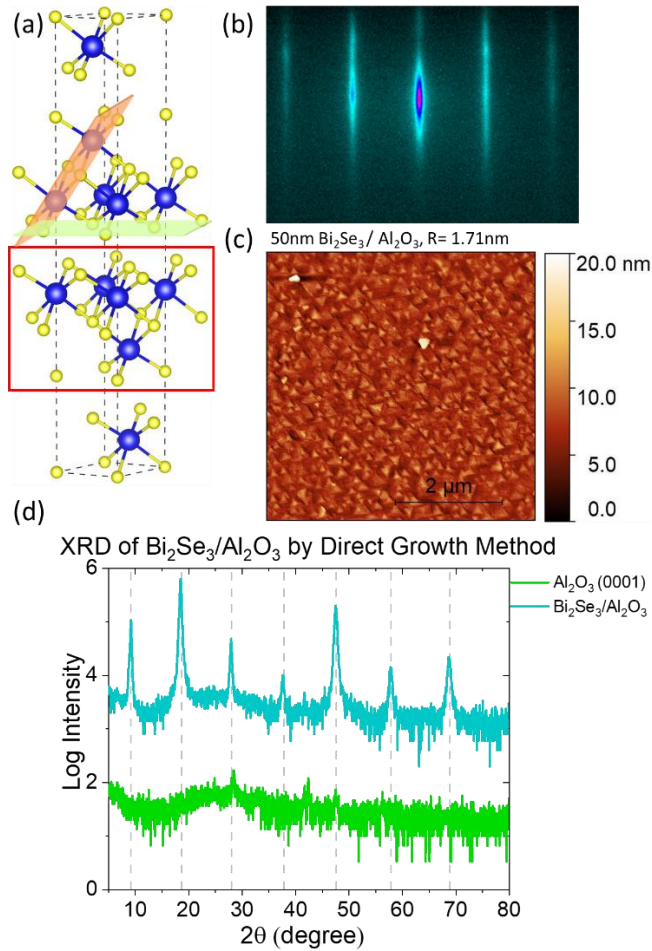

**Figure 1.** (a) Crystal structure of quintuple layer of Bi$_2$Se$_3$, yellow atoms: selenium, blue atoms: bismuth. The red box marks one quintuple layer (QL) of Bi$_2$Se$_3$. The green plane indicates the (0001) plane while the red indicates the (10$\bar{1}$5) plane, which has the highest intensity in powder XRD pattern. (b-d) are structural characterization of sample A-1: 50nm Bi$_2$Se$_3$ on sapphire via direct co-deposition: (b) RHEED pattern (c) AFM scan over a 5 $\mu$m×5 $\mu$m area. The root mean square roughness R is marked above the image. (d) XRD curves for the sapphire substrate (green), and Bi$_2$Se$_3$ (sample A-1 in blue) directly grown on top. The light grey dashed lines mark the peaks associated with the Bi$_2$Se$_3$ (0003) peak series. The absence of the sapphire peak in the curves is due to the peak width being narrower than (or comparable to) the XRD scanning step. Curves are offset by 2 for clarity.

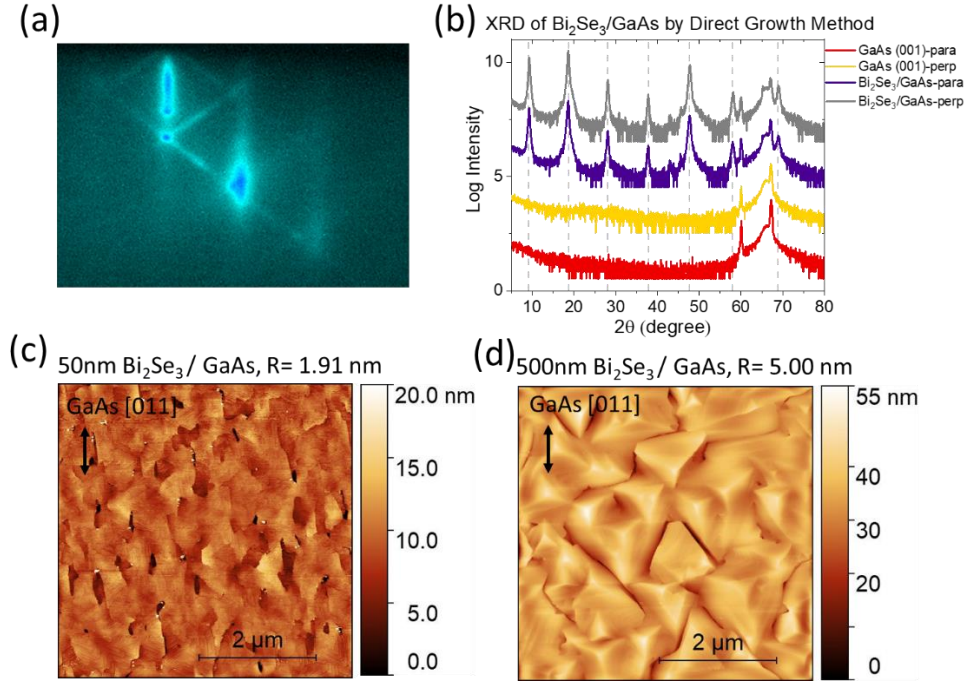

Figure 2. Structural characterization of sample A-2: 50nm $Bi_2Se_3$ on GaAs via direct co-deposition: (a) RHEED pattern composed of 2×4 GaAs reconstruction and chevron lines from the $Bi_2Se_3$ ($10\bar{1}5$) orientation at the early stages of the growth. (b) XRD curves for the GaAs substrate (yellow, red), and $Bi_2Se_3$ (sample A-2 in grey and violet) directly grown on top of them. Due to the anisotropy of the GaAs surface after deoxidation, we take XRD scans both parallel and perpendicular to the [$0\bar{1}1$] directions of GaAs (100) wafer. Curves are offset by 2 for clarity. (c) AFM image of 50nm $Bi_2Se_3$ on GaAs. (d) AFM image of 500nm $Bi_2Se_3$ on GaAs. Black arrows mark the GaAs [011] axis.

### 3A. Direct co-deposition

In this method (Strategy A), selenium, bismuth, and indium (if used) fluxes are supplied simultaneously. RHEED is checked in the first 10min and at the end of the growth to monitor the film quality *in situ*. On sapphire, the substrate temperature is kept at 325°C and on GaAs, the substrate is kept at 375°C. Co-deposition is the simplest way to grow these materials and is a commonly-used technique. We will use films grown this way as the baseline against which we will compare all other growth methods.

i. **On sapphire (Sample A-1)**

For $Bi_2Se_3$ grown on sapphire substrates, the RHEED image changes from the sapphire pattern into polycrystalline rings within 0.5QL after opening the bismuth and selenium shutters. As the growth proceeds, streaks start to appear when the film thickness reaches 2-3QL, and the ring features disappear when the film thickness reaches 5QL, resulting in what we see in Figure 1(b). Figure 1(c) shows the AFM picture of the $Bi_2Se_3$ on sapphire. The surface roughness is characterized by root mean square roughness R, which is given above the image. On sapphire, we see in Figure 1(c) the typical triangular pyramidal domains of $Bi_2Se_3$ with a background of complete coalesced layers. Furthermore, we can see that some of these pyramids are rotated 60° relative to the others, indicating the formation of twin or antiphase domains, which we will discuss in detail in section 3D. The excellent crystallinity and single orientation of the $Bi_2Se_3$ film grown using Strategy A is further confirmed via wide range XRD coupled $2\theta$ scans, as shown in Figure 1(d). We use four-index notation (hkil) for crystals that have a conventional hexagonal unit cell such as sapphire and $Bi_2Se_3$. GaAs with a cubic zinc blende structure uses the three-index notation (hkl). For the $Bi_2Se_3$ grown on the sapphire substrate, we see only the typical $Bi_2Se_3$ (0003) peaks as expected, shown in Figure 1(d) with the blue curve. The sapphire XRD is also shown for comparison in the green curve.

ii. **On GaAs (Sample A-2)**

For the growth on GaAs substrates, we start with the RHEED pattern corresponding to the 2×4 surface reconstruction of the GaAs (001) surface after deoxidation under selenium flux. As the growth begins, within 0.5QL the RHEED pattern evolves into the RHEED pattern shown in Figure 2(a) at specific rotation angles. These chevron lines between the streaks indicate that the $Bi_2Se_3$ film is exhibiting multiple orientations and/or three-dimensional growth and that these features are aligned to the substrate. These features in the RHEED pattern have been confirmed to be needle-like aggregates in the AFM images (Figure 9(a) shows a very similar AFM though it is for BIS on GaAs). AFM images and a detailed discussion of how growth parameters control the growth orientation is included in ref.[64].

Because of the in-plane anisotropy we see in the RHEED, we measured the XRD pattern in two orthogonal directions: in Figure 2(b), the red (yellow) curve shows the XRD pattern for the bare GaAs substrate with the x-ray beam incident plane parallel (perpendicular) to the GaAs $[0\bar{1}1]$ axis. The $Bi_2Se_3$ film grown on GaAs is characterized similarly, as shown in the violet (parallel to GaAs $[0\bar{1}1]$) and gray (perpendicular to GaAs $[0\bar{1}1]$) curves. As pointed out in Ref[64,65], an extra peak at 29.5° appears if the unwanted $Bi_2Se_3$ $(10\bar{1}5)$ orientation is present as a non-negligible fraction of the film. Because of the $(10\bar{1}5)$ preferential alignment with the GaAs [011] axis, the XRD diffraction peak is more prominent when the x-ray beam is parallel to the GaAs $[0\bar{1}1]$ axis. Here in both directions, we only see the diffraction peaks from the GaAs substrate and the $Bi_2Se_3$ (0003) peaks (positions marked by vertical gray dashed lines), indicating that the majority of the $Bi_2Se_3$ film is oriented in the [0001] orientation. As shown in AFM in Figure 2(c), we can see that for the same thicknesses, the film on GaAs has comparable roughness to the film on sapphire. However, the surface morphology is very different. We do not see the triangular shaped domains at the same size scale. This can be attributed to the fact that the film is grown at a higher temperature, so the domains are larger and the topmost layer is better coalesced. In addition, the 2×4 GaAs surface reconstruction may have inhibited the triangular domain formation but promoted the elongated domain formation due to preferential bismuth diffusion along [011] GaAs axes. For films with a thickness of 500nm, large triangular pyramids are present and the relative roughness (r=R/d, roughness R divided by film thickness d) has gone down (Fig. 2(d)).

(a) 10nm $Bi_2Se_3/Al_2O_3$, direct, R=1.72 nm

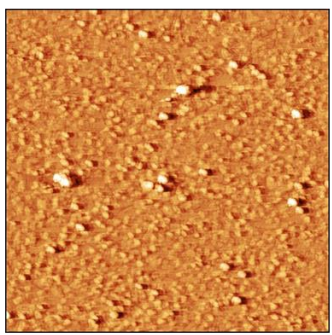

(b) 10nm $Bi_2Se_3/Al_2O_3$, grow/anneal, R=1.15 nm

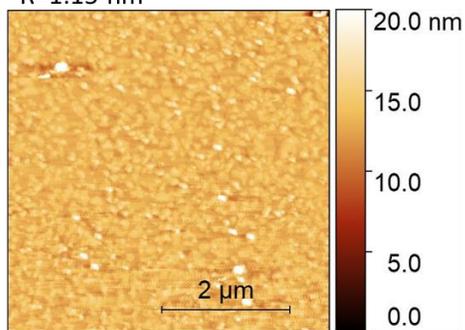

(c)

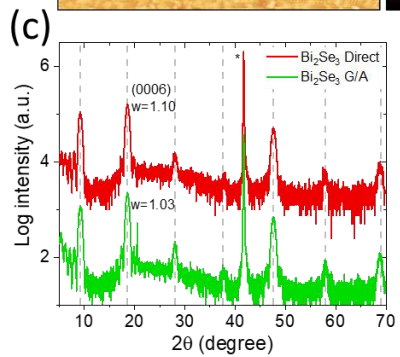

**Figure 3.** Characterization of $Bi_2Se_3$ grown on sapphire substrates with strategy B: grow-anneal deposition. AFM of 10nm $Bi_2Se_3$ grown via (a) direct deposition and (b) grow/anneal deposition at the same color scale. (c) XRD scans for 10nm $Bi_2Se_3$ via direct co-deposition (red) or via grow/anneal deposition (green). The asterisk marks the substrate peak while the grey dashed lines marked the (0003) series peaks. The numbers near each scan indicate the full width at half maximum of the (0006) peak. Curves are offset by 2 for clarity.

**3B. Grow-anneal**

For this growth technique (Strategy B), all the growth parameters are the same as strategy A, except that we repeatedly open the bismuth shutter for one minute to grow the film and close the shutter for ten minutes to anneal. The accumulated bismuth shutter opening time equals the time for direct co-deposition growth. These types of grow-anneal cycles are commonly used in conventional semiconductor MBE to improve film quality. The anneal portion of the cycle is intended to improve adatom mobility, which we hypothesized would result in larger domains and a higher electron mobility for these materials.

**On sapphire (Sample B-1: $Bi_2Se_3$)**

In Figure 3, we present the AFM and XRD scans of $Bi_2Se_3$ grown via the grow-anneal strategy and compare them to the direct growth. For $Bi_2Se_3$ on sapphire, the annealing time helps decrease the surface roughness from R=1.72nm to R=1.15nm. However, the XRD scans do not show significant improvement in the film crystallinity as determined by the full width at half maximum (FWHM) of the peaks, as shown in Figure 3(c). In Hall measurements, we observed nearly the same carrier density and a small improvement in mobility for the grow-anneal sample.

**On GaAs**

The grow-anneal strategy shows a significant effect on the morphology of $Bi_2Se_3$ on GaAs[64]. It was shown that more annealing time favors the $(10\bar{1}5)$ orientation of $Bi_2Se_3$ flakes while the size of the flakes shrinks. Again, the bismuth adatoms have long diffusion lengths along the GaAs [011] axis, which promotes the formation of the $(10\bar{1}5)$ orientation with more annealing time, rather than the desired (0001) orientation. For details, please see ref.[64].

**3C. Two-step growth**

In this growth strategy, for both sapphire and GaAs, the first five QLs of $Bi_2Se_3$ are grown using direct co-deposition at a lower temperature of 325°C to wet the substrate, and then the film is heated to 425°C under a selenium flux. The rest of the film is deposited at this temperature via co-deposition. Similar techniques have been tried by other groups, though with different temperatures used for the low- and high-temperature steps[66–68]. As noted above, it can be difficult to accurately compare temperatures from one system to the next without the use of an objective temperature standard, so it is somewhat challenging to compare our temperatures to those reported by other groups.

We hypothesized that film growth at a higher substrate temperature would enhance adatom mobility, increase domain size, and improve mobility. Films cannot be grown directly at these high substrate temperatures due to a reduced sticking coefficient. Instead, during the low temperature growth step, $Bi_2Se_3$ islands form and coalesce to form larger domains. The in-plane covalent bonds formed during the coalescence phase are stable up to the second step temperature, which is slightly below the thermal decomposition point.

    i. **On sapphire (Sample C-1 through C-4)**

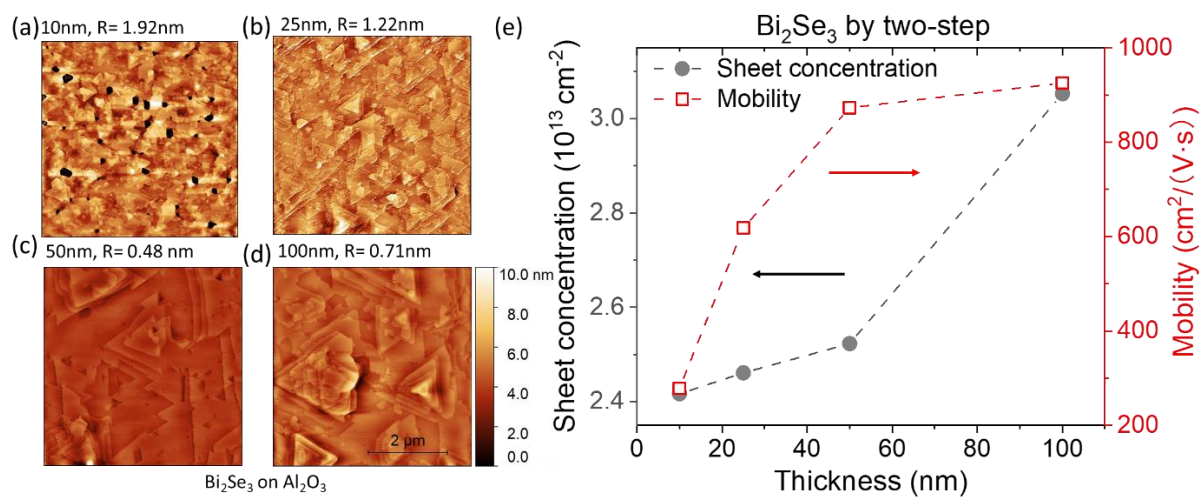

**Figure 4.** Thickness dependence study of Bi$_2$Se$_3$ grown on sapphire via two-step strategy with growth rate 0.4nm/min. (a-d) AFM scans of 10nm, 25nm, 50nm, 100nm films, respectively. (e) Sheet carrier concentration (grey circles) and mobility (red empty squares) as a function of film thickness. The dashed lines serve as visual guides.

In Figure 4, we present a thickness dependence study of the two-step growth to understand how the film evolves. For all the films, the RHEED images show very narrow and bright streaks at the end of the growth, which indicates good crystallinity and smooth surfaces for these films. Comparing Figure 4(c) and Figure 1(c), we can see that for the same thickness, the domain size is larger in the sample grown with the two-step strategy than in the sample grown with direct growth (sample A-1). In addition, the surface roughness has dropped to R=0.48nm, significantly lower than in the co-deposited films. The room temperature Hall measurement shows a 20% decrease in carrier concentration and nearly 50% increase in carrier mobility.

As the film thickness increases, shown in Fig. 4(a-d), the domain size increases and the relative surface roughness decreases monotonically. This is very similar to conventional III-V material MBE growth where further growth causes the surface to get flatter. However, for many chalcogenides, additional growth causes roughening because the vdW interaction means that addition growth results in additional domain nucleation. We attribute this contradictory observation to the greatly improved adatom surface mobility which causes more layer-by-layer growth when using the two-step growth strategy. The slight increase in absolute surface roughness of 100nm film is likely from thermal roughening. $Bi_2Se_3$ has a narrow bandgap. As the film gets thicker, more of the thermal radiation from the heater gets absorbed, so the actual surface temperature rises. This effect is significant in InAs (bandgap $E_g$=0.36eV) deposited on GaAs ($E_g$=1.42eV): the film temperature experiences a sharp increase from 370°C to nearly 520°C within 200nm of InAs deposition[69]. Considering that the bandgap of sapphire is even larger at $E_g$=9.1eV[70], far larger than that of the $Bi_2Se_3$ ($E_g$=0.33eV), the radiation from the heater that is between the two energies may be absorbed by the $Bi_2Se_3$ film, potentially leading to a large temperature as in the InAs-GaAs system. This effect can also explain the dramatic increase in sheet carrier density for the 100nm film as shown in Figure 4(e): higher temperatures result in more selenium vacancies leading to higher n-type doping. For films thinner than 50nm, the sheet density slowly increases with thickness, indicating a relatively small bulk carrier concentration and that the 2D conduction channel at the surface dominates. We also note that for the 10nm film, the carrier mobility is 60% larger than the film grown via direct growth, while the surface roughness is comparable. The improvement in mobility can be explained by the larger domain size leading to fewer twin defects. As the film thickness increases, the mobility increases due to the reduction in surface defect scattering. The mobility approaches a saturation value of just over 900 $cm^2$/(V • s).

  ii. **On GaAs (Sample C-5)**

For $Bi_2Se_3$ grown on GaAs, we see a worse morphology, as shown in Figure 5. The domains are more fractal-shaped and deep trenches are present. The surface roughness has increased compared to direct growth. We attribute this to the anisotropic bismuth migration on the GaAs substrate at high temperature described earlier and preferential incorporation of adatoms on top rather than the edges of the domain. From the Hall measurements (sample C-5), we see a reduction in carrier density due to the improvement in crystallinity, but the deep trenches have reduced the mobility.

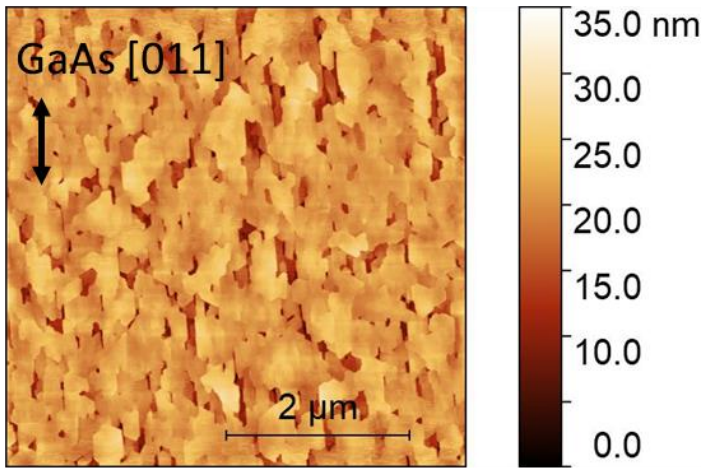

Figure 5. AFM scan of 50nm $Bi_2Se_3$ grown on GaAs with deposition rate 0.74nm/min.

**3D. Predose growth**

The fourth strategy we tried is pre-dose treatment of the substrate. It has been shown that by exposing the substrate to the material flux above the thermal decomposition temperature, single crystal (111)-oriented PbSe can be epitaxially grown on nearly lattice-matched InAs (111) A templates[71]. Twin formation has been suppressed by a pre-dose of the epitaxy material. A similar method of pre-dosing the substrate with $Bi_2Se_3$ has been tried on sapphire: a bismuth predose of 2 minutes at a low temperature was used followed by growth of a buffer layer of $Bi_2Se_3$, which is then re-evaporated in a flash heat up process. After the pretreatment, the actual $Bi_2Se_3$ is grown via a two-step growth strategy[36]. The mechanism behind the twin suppression in films has been attributed to realignment of domains during evaporation and formation of bismuth terminated surface (e.g. Bi-O bonds at the interface)[36].

In our samples, before the film growth, the substrate is set at a temperature higher than the $Bi_2Se_3$ thermal decomposition point 435°C, and it is exposed to bismuth and selenium fluxes (the same as for the growth) for 5 minutes. After the predose, the substrate is cooled to 325°C for sapphire and 375°C for GaAs for growth by direct co-deposition as in samples A1 and A2. With the predose step, we aimed to passivate the substrate with both bismuth and selenium atoms so that the tiny domains can align and serve as seeds for twin free films.

    i.    **On sapphire (Samples D1-D3)**

In Figure 6, we present AFM images of $Bi_2Se_3$ grown on sapphire with pre-growth exposure at substrate temperatures of 450°C (Fig. 6(a)), 440°C (Fig. 6(b)), and 425°C (Fig. 6(c)). During the 5 minute exposure process, the first two temperatures do not show any change in the RHEED pattern, while the last one starts to show polycrystalline rings in the RHEED pattern after approximately 1 minute of exposure. After the exposure, we bring the substrate temperature back to 325°C for co-deposition growth. From these AFM images, we can see that this type of pre-dose growth does not improve the film surface significantly when compared to the direct growth. The sample grown with pre-dose at 425°C shows a relatively high roughness of R=4.24nm. This is likely attributable to poor surface wetting of the $Bi_2Se_3$ that starts at 425°C, which was observed in RHEED. The poor surface wetting at high temperature also explains why we have to create seed layer at lower temperature in strategy C, the two step growth. The rough morphology of the wetting layer influenced films as thick as 50nm. In the other two samples, the substrate wetting starts at the growth temperature at 325°C. However, these three films do not show much significant transport improvement as expected, from the Hall measurements data.

Twinning is often observed in hexagonal crystal thin film growth. Wherever the twin domains coalesce, twin boundaries form and act as one of the major scattering sites for the surface state electrons. Therefore, the suppression of twin domains is another priority for $Bi_2Se_3$ thin films and an indicator of film quality. In order to quantitatively evaluate the twin domain improvement, we did a ψ scan of the $(10\bar{1}5)$ plane of $Bi_2Se_3$, which has the highest intensity in powder XRD diffraction. Rhombohedral $Bi_2Se_3$ has a three-fold symmetry, so the $(10\bar{1}5)$ plane (indicated in red in Figure 1(a)) should only show three equally spaced peaks in a complete 360° ψ scan. However, if the film exhibits twin domains rotated 60° relative to each other, a six-fold symmetry would be expected. A suppression of twin domain formation can be judged by the relative intensity of the $(10\bar{1}5)$ planes from one domain orientation compared to the other. In Figure 6(d), we show the ψ scan of $Bi_2Se_3$ on sapphire grown by the predose growth strategy. Six peaks are present in each sample, which can be attributed to the two twin orientations. Each orientation has peaks separated by 120°. An average peak intensity (or peak area) for each orientation set is calculated, from which we estimate an approximate twin domain ratio. The film grown with the 440°C predose treatment shows the best suppression of twin domains. This is because when the temperature is slightly above the thermal decomposition temperature, even though we do not deposit any material that can be observed with RHEED, it is likely that the sapphire surface is being altered by small particles of bismuth or bismuth selenides (not necessarily stoichiometric). This is especially true for the active sites, for example, at the step edges and/or dislocation ends. Because of the high temperature, the small domains can rotate and form larger, single oriented domains. If the temperature is far higher than thermal decomposition point, the predose flux may not interact with the substrate at all. If the temperature is lower than thermal decomposition point, the predose flux will give rise to the normal twin domain formation since the substrate temperature is not hot enough for the small domains to surpass the rotational energy barrier.

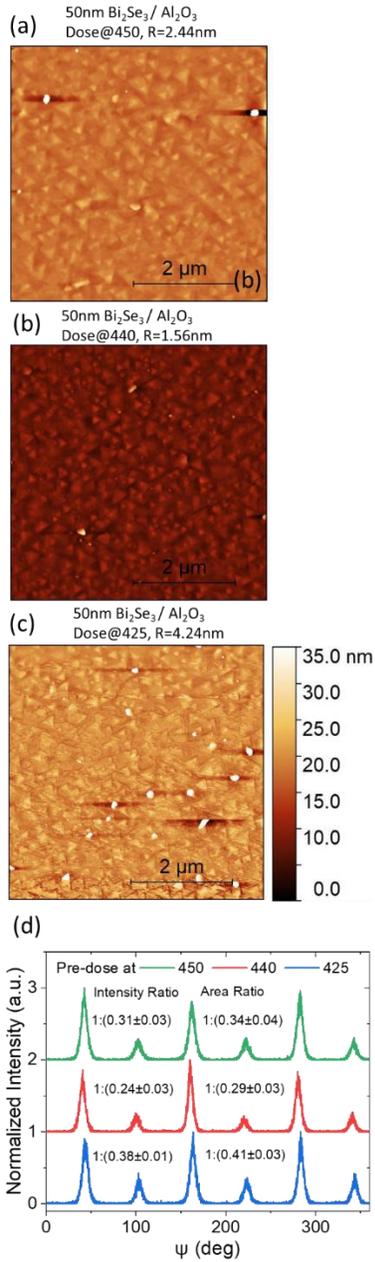

**Figure 6.** (a-c) AFM of the 50nm Bi$_2$Se$_3$ grown via pre-dose strategy on sapphire, pre-dose temperature at (a) 450°C (b) 440°C (c) 425°C. (d) ψ scan of the predose samples on sapphire: the average peak intensities (peak area) are calculated from Lorentz fit. From the relative ratio of the two orientation sets, the sample with 440°C predose treatment shows the most suppressed, with 80% in one orientation. The curves are offset by 1 for comparison.

### ii.   On GaAs (Samples D4-D6)

We applied the same predose method to GaAs substrate but observed very different behavior. Figure 7(a-c) presents the AFM scans for $Bi_2Se_3$ films grown on GaAs (001) with the predose at temperatures from 450°C to 425°C. The surface roughness and morphology of the three films are similar except that we see elongated domains for samples exposed at 425°C. This is consistent with our theory that the $Bi_2Se_3$ starts to wet the GaAs substrate at 425°C, and because the high growth temperature favors the $(10\bar{1}5)$ orientation growth, the wetting layer formed at this temperature composes a large fraction of $(10\bar{1}5)$ orientation, which promotes anisotropic in-plane epitaxy at 375°C. For all three samples, when the growth starts at 375°C, the (0003) orientation soon outgrows the $(10\bar{1}5)$ plane and buries these unwanted domains. This is why we do not observe the $(10\bar{1}5)$ diffraction at 29.5° in wide angle XRD scans, as shown in Figure 7(d). As show in Figure 7(e), the ψ scans of this batch of samples are dissimilar from the films on sapphire. We observe multiple features including (1) narrow and high intensity peaks marked by asterisks from GaAs substrate (111) planes every 90°; (2) the $(10\bar{1}5)$ planes of (0003) oriented domain marked by grey dashed lines, part of which overlap with the GaAs (111) peaks; (3) extra small peaks marked by pound signs show up every 60°, which are from the (0003) plane diffraction of the $(10\bar{1}5)$ oriented domains. Although the overlapping of the $(10\bar{1}5)$ peak and substrate peak makes it hard to perform an analysis similar to Figure 6(d), from the peaks at 163° and 223°, the twin domain ratio is basically 1:1 for all three samples, which indicates that this pre-exposure strategy does not reduce the twin domain formation on GaAs.

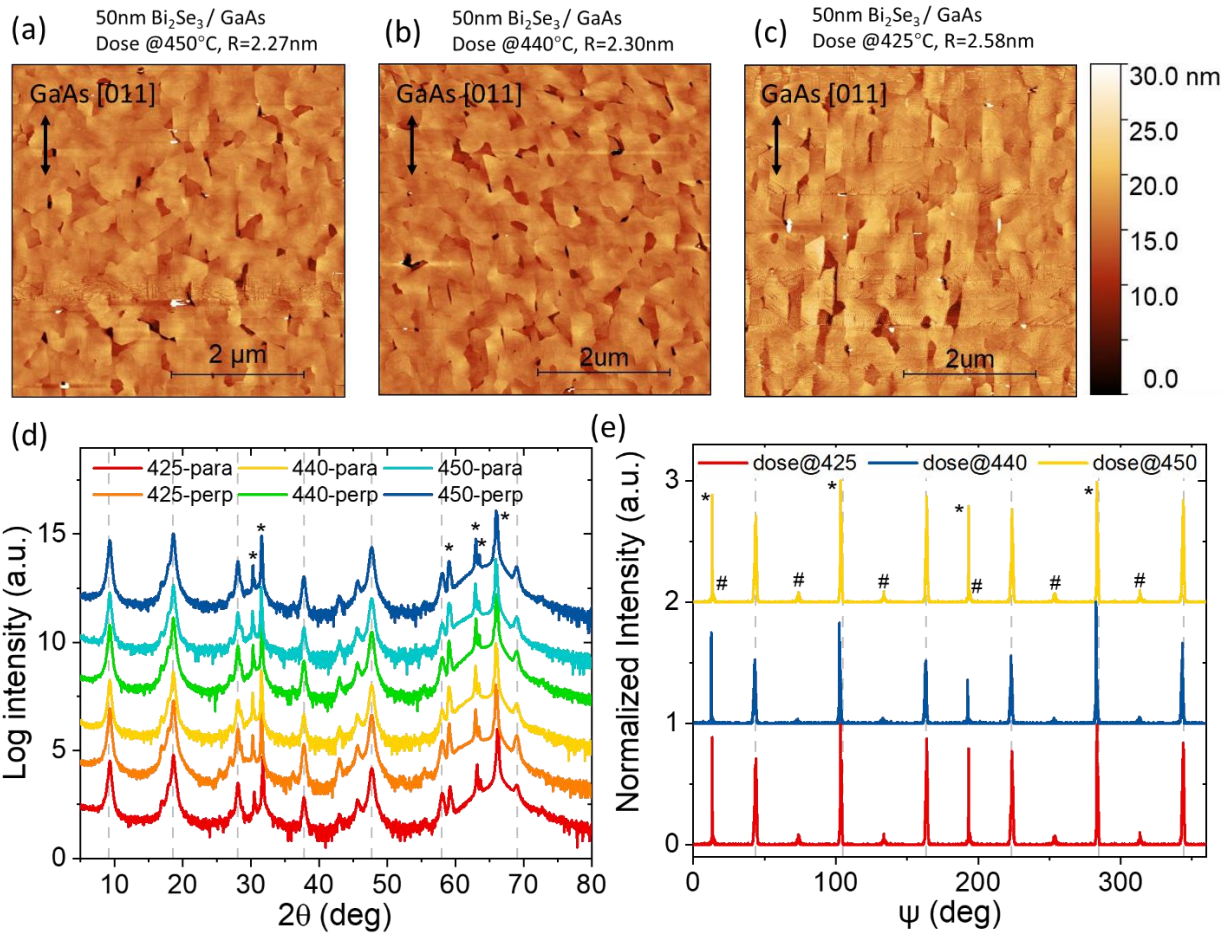

**Figure 7.** (a-c) AFM of the 50nm Bi$_2$Se$_3$ grown via pre-dose strategy on GaAs (001) plane, pre-dose temperature at (d) 450°C (e) 440°C (f) 425°C. (d) XRD scans of Bi$_2$Se$_3$ predose growth on GaAs, with x-ray incident plane parallel (para) or perpendicular to [0$\bar{1}$1] axis: Grey dashed lines marked the (0003) planes diffraction, while the asterisks mark the diffraction peaks from the substrate. The curves are offset by 2 for comparison. (e) ψ scan of the predose samples on GaAs: grey dashed lines mark the peaks from Bi$_2$Se$_3$ (10$\bar{1}$5) planes of the (0003) oriented twin domains. The asterisks mark the diffraction peak from GaAs (111) plane at the same tilt angle of 58° from the film normal. The pound signs mark the small peaks from the (0003) planes of the (10$\bar{1}$5) oriented domains that are buried underneath the film. The curves are offset by 1 for comparison.

**3e. Se sacrificial layer growth**

The final strategy we tried is the use of a selenium sacrificial layer. In this strategy, we first deposit an elemental selenium layer at a low temperature (near room temperature) and then grow 1-2QLs of $Bi_2Se_3$ on top. We then heat up the substrate to the direct growth temperature of $Bi_2Se_3$. During the heating, the extra elemental selenium evaporates, and we hypothesize that the small domains of $Bi_2Se_3$ will coalesce and form a more complete layer in the (0003) orientation. This method has been tried on rare earth iron garnets substrates like thulium iron garnet[72].

We expected that this particular strategy would be most beneficial on GaAs rather than on sapphire. Because of the un-passivated dangling bonds after deoxidation, the in-plane $Bi_2Se_3$ vdW epitaxial growth ((0001) orientation) on GaAs is in competition with the $Bi_2Se_3$ out-of-plane traditional epitaxy growth ((10$\bar{1}$5) orientation), especially at the early stages of growth as observed in RHEED. Although the (0001) orientation will dominate as the growth proceeds, the existence of the (10$\bar{1}$5) flakes will affect the overall film crystallinity and electronic transport properties. One way to suppress the origin of the (10$\bar{1}$5) growth is to passivate the GaAs surface with the elemental selenium layer. The selenium layer may act as separation material for the $Bi_2Se_3$ and the GaAs substrate and weaken the $Bi_2Se_3$-GaAs interaction in the early growth stages. We hypothesize that after selenium re-evaporation, a more complete layers of $Bi_2Se_3$ will remain in the (0003) orientation and bonded to GaAs via vdW forces. We did not expect a significant improvement for growth on sapphire using this method.

    i.      **On sapphire (Sample E-1)**

To begin, we first investigate the morphology and crystallinity of an elemental selenium layer deposited on sapphire at <80°C using the same selenium flux that we use for the $Bi_2Se_3$ growth. The deposition lasted for 30min. The RHEED pattern comprises of rings when the growth begins and becomes dimmer as growth proceeds. The AFM scan presented in Figure 8(a) indicates the formation of amorphous selenium droplets on the surface, which can be explained by the self-passivation of the sapphire substrate and the high surface tension of selenium. The XRD scan in Figure 8(c) also confirms that the selenium layer deposited directly on sapphire is amorphous. Profilometry indicates the thickness of the selenium layer is around 150nm, giving a selenium deposition rate of about 5nm/min with current flux.

We then tried the selenium sacrificial layer strategy with $Bi_2Se_3$ on sapphire. After a 5nm selenium passivation layer deposition followed by 2QLs of $Bi_2Se_3$, the film is heated directly to 325°C for 3QLs deposition to form the wetting layer. Then the film is heated up further to 425°C for the rest of the film. However, neither the film surface morphology (shown in Figure 8(b)) nor the electronic transport properties (sample E-1) are improved compared to the two-step method without the selenium sacrificial layer (sample C-3).

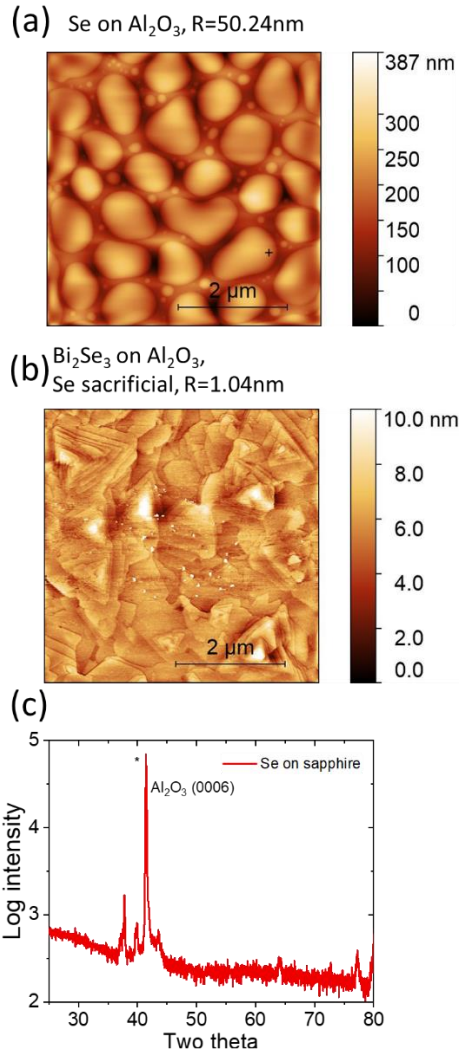

Figure 8. AFM of: (a) 150nm selenium layer deposition on sapphire. (b) $Bi_2Se_3$ on sapphire, grown via selenium sacrificial layer strategy, combined with the two step growth method. (c) XRD pattern of elemental selenium layer deposited on sapphire. Only the sapphire diffraction peak is visible.

### ii. On GaAs (Sample E-2)

An elemental selenium layer deposited on GaAs, on the other hand, looks crystalline in both the AFM image, shown in Figure 9(a), and the XRD measurement, shown in Figure 9(c). Crystalline selenium also has a trigonal unit cell comprised of selenium atomic chains. In the XRD scan, we saw a small peak corresponding to the (0003) plane of selenium and a large peak corresponding to the ($2\bar{1}\bar{1}0$) plane. This indicates that the selenium atomic chains are aligned with the GaAs [011] direction.

For $Bi_2Se_3$ grown on GaAs using this strategy, the RHEED pattern shows a polycrystalline ring formed during the deposition of approximately 7nm of selenium and 1-2QLs of $Bi_2Se_3$ on top of the 2×4 GaAs RHEED background. When the substrate reaches 190°C, we resume the selenium supply to prevent selenium outgassing from the $Bi_2Se_3$ layer. When the substrate temperature reaches 250-265°C, the RHEED image shows a dramatic change: the GaAs 2×4 reconstruction pattern disappears, leaving only polycrystalline rings. We think that at this temperature, most of the selenium passivation layer (composed of selenium atoms that are not directly connected to gallium) has evaporated. As the co-deposition proceeds at 350°C, the RHEED streaks associated with the $Bi_2Se_3$ film appear and brighten, while the polycrystalline ring continues until the end of the growth. The RHEED patterns indicate that the film has both 3D growth and layer-by-layer growth of $Bi_2Se_3$ at this temperature, which is further confirmed in Figure 9(b) with an AFM scan. Similar triangular or hexagonal shaped nanocolumns have been observed before in $Bi_2Se_3$ grown on a BIS seed layer on GaAs or on sapphire at high temperatures, without applying the selenium sacrificial layer[63]. These nanocolumn features roughen the surface, reducing the carrier mobility and increasing the carrier density. However, we do not see either the ($10\bar{1}5$) orientation or the preferred growth along the GaAs [011] direction. In other words, by applying the selenium sacrificial layer, we suppressed the ($10\bar{1}5$) plane growth and we promoted the isotropic in-plane growth. It is suggested that to suppress the nanocolumn growth, a reduction of selenium overpressure and lower substrate temperatures are needed[63].

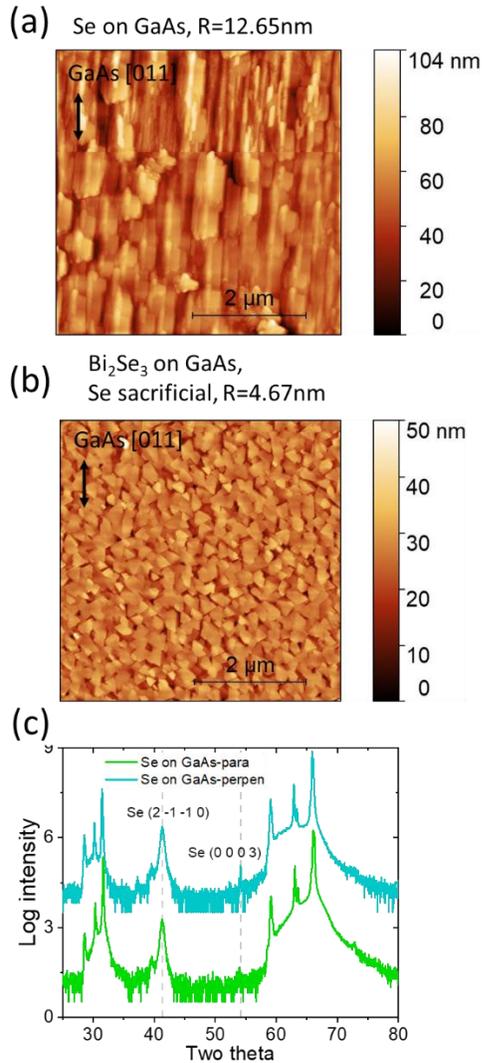

**Figure 9.** (a) AFM image of selenium deposition on GaAs substrate at 65°C, crystalline selenium domains are aligned in GaAs [011] direction. (b) AFM image of Bi$_2$Se$_3$ grown on GaAs at 350°C. (c) XRD pattern of elemental selenium layer deposit on sapphire (red) and on GaAs, in direction parallel (green) and perpendicular (blue) to GaAs [0$\bar{1}$1] axis. The curves are offset in log intensity by 3 for comparison.

### 3f. BIS growth on GaAs

As noted in a variety of previous papers, a BIS buffer layer can be used to improve the growth of Bi$_2$Se$_3$[24,57,73]. Although there has already been work exploring how the selenium overpressure and bismuth to indium ratio affects the film morphology[74], we still lack a growth recipe for a smooth, single orientation BIS film on GaAs. We will now describe our efforts to obtain such a recipe using the aforementioned growth methods to show their applicability to general van der Waals material no matter what band structure they have.

### i. Two-step growth (Sample F-1)

The first method we tried was the two-step method, where we first grow 10nm of BIS at a lower temperature of 325°C and then grow the rest of the layer at 425°C. This is sample F-1. The RHEED images show chevron lines throughout the entire growth, which indicates growth of the (10$\bar{1}$5) orientation. The needle-like domains observed in the AFM image, as highlighted by the yellow box of Figure 10(a), have further confirmed the growth of this orientation. However, the diffraction peaks shown in XRD scans taken parallel and perpendicular to the GaAs [011] direction are almost identical. The BIS (0003) series peaks are marked with grey dashed lines, while the extra peaks are marked with black dashed lines and pound symbols. We note that peak #1 and peak #2 appear exclusively in sample F-1. These two peaks are attributed to the cubic structured κ-phase In$_2$Se$_3$ (003) and (005) plane, which indicate possible phase segregation happening in the film. Peak #3 can be attributed to κ-phase In$_2$Se$_3$ (006) plane and it is present in all the films except sample F-4. We did not see the (10$\bar{1}$5) orientation peak, which we expect to be at around 29.5° when the beam is parallel to the GaAs [0$\bar{1}$1] axis. This may be caused by the relatively small fraction of (10$\bar{1}$5) orientation as compared to the regular (0003) orientation. In addition, we do not see the prolonged fractal like in-plane domain structure that we saw in Bi$_2$Se$_3$ grown by the two-step method on GaAs (Figure 5). The added indium atoms may have inhibited the preferred migration of bismuth atoms along the GaAs [011] axis.

### ii. Selenium passivation layer (Sample F-2)

To suppress the $(10\bar{1}5)$ orientation, we switched to the selenium passivation layer strategy to grow sample F-2. We grew an approximately 10nm-thick elemental selenium sacrificial layer followed by 2QL of $Bi_2Se_3$ at 65°C. We then heated the sample to 350°C and grew the additional 3QLs of $Bi_2Se_3$ and 5QLs of $In_2Se_3$ to form a seed layer, followed by deposition of the rest of the BIS layer via co-deposition at the same temperature. The RHEED patters were very similar to what we observed for sample E-2: the RHEED pattern for selenium and the GaAs 2×4 reconstruction disappeared and only the polycrystalline ring remains after heating the sample to 250-265°C. Streaks start to form in RHEED when the co-deposition starts, but the rings remain until the end of growth. A columnar surface morphology is confirmed by AFM scans, shown in Figure 10(b). No $(10\bar{1}5)$ $Bi_2Se_3$ orientation is observed in the XRD traces (green and yellow curves in Figure 10(e). The nanocolumn formation is likely caused by bismuth and indium adatom incorporation being more energetically favorable on the tip of the column rather than on the side walls. In addition, the selenium passivation layer on the GaAs substrate may cause a rough surface to start with, which has led to a rough $Bi_2Se_3$ seed layer, and thus inhibited the adatom mobility along the surface, resulting in atoms diffusing along the side wall of the nanocolumns instead.

### iii. Direct growth (Sample F-3 through F-5)

Sample F-3 was grown via direct growth at 375°C and we again observe the formation of nanocolumns. We see small $(10\bar{1}5)$ orientation needles between the hexagonal domains, indicated by the yellow box in Figure 10(c). We therefore hypothesized that the column-like growth will be suppressed with a lower substrate temperature and therefore lower bismuth adatom mobility. Sample F-4 was grown at 300°C by direct deposition. A streaky RHEED patterned formed after the seed layer and continued until the end of the growth. The film grown directly at a low substrate temperature is thus the smoothest we were able to obtain, and we only see the (0003) family of BIS peaks and the substrate peaks in the XRD scan. The extremely tall white dots in Figure 10(d) are dust on the sample surface. Finally, we used the recipe for sample F-4 to grow a BIS buffer layer on GaAs and grow a 50nm $Bi_2Se_3$ on top at 300°C (sample F-5). The sheet density has reduced in half compared to $Bi_2Se_3$ directly grown on GaAs and carrier mobility shows a nearly 60% improvement compared to sample A-2.

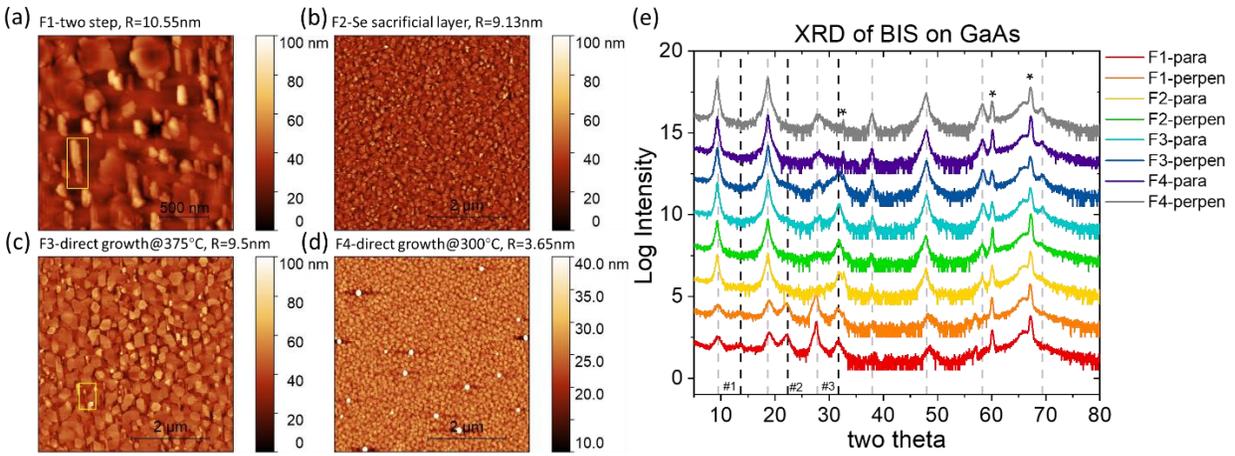

**Figure 10.** (a-d) AFM scans of BIS growth on GaAs substrate using: (a) two step growth strategy, (b) selenium sacrificial layer strategy, (c) direct growth strategy with substrate at 375°C, (d) direct growth strategy with substrate at 300°C. (e) XRD scans of sample F-1 to F-4 with x-ray beam incident plane parallel or perpendicular to the [0$\bar{1}$1] axis of GaAs substrates. The grey dashed lines mark the (0003) serial peaks; asterisks mark the substrate peaks; the black dashed lines mark the extra unidentified peaks that are present in some of the samples, the lines only serve as guides to the eyes. Curves are offset by 2.

**Table 1**. Room temperature Hall effect data of Bi$_2$Se$_3$ growth on Sapphire

| Sample | Growth Description | Carrier density ($10^{13}$ cm$^{-2}$) | Mobility (cm$^2$/(V·s)) |
|---|---|---|---|
| A-1 | 50nm: direct growth, 0.74nm/min, substrate at 325 °C | 3.11±0.02 | 589.09±0.38 |
| B-1 | 10nm: 1 min grow/ 10 min anneal | 2.81±0.01 | 188.46±1.03 |
| B-2 | 10nm: direct growth | 2.90±0.04 | 167.91±1.95 |
| C-1 | 10nm: 5nm at 325°C, 5nm at 425°C, 0.4nm/min | 2.41±0.01 | 278.16±3.61 |
| C-2 | 25nm: 5nm at 325°C, 20nm at 425°C, 0.4nm/min | 2.46±0.02 | 618.23±2.26 |
| C-3 | 50nm: 5nm at 325°C, 45nm at 425°C, 0.4nm/min | 2.52±0.00 | 873.06±0.17 |
| C-4 | 100nm: 5nm at 325°C, 95nm at 425°C, 0.4nm/min | 3.05±0.01 | 925.66±1.56 |
| D-1 | 50nm: pre-dose at 450°C, growth at 325°C | 2.91±0.04 | 573.4±8.18 |
| D-2 | 50nm: pre-dose at 440°C, growth at 325°C | 3.39±0.02 | 589.09±0.38 |
| D-3 | 50nm: pre-dose at 425°C, growth at 325°C | 3.00±0.02 | 608.2±1.89 |
| E-1 | 5nm Se sacrificial + 50nm Bi$_2$Se$_3$ two step as C-4 | 2.68±0.04 | 811.31±1.35 |

**Table 2.** Room temperature Hall effect data of Bi$_2$Se$_3$ growth on GaAs. A and B means we cleaved out two 1cm×1cm square chips from a quarter of 2-inch film and we measure them respectively.

| Sample | Growth Description | Carrier density ($10^{13}$ cm$^{-2}$) | Mobility (cm$^2$/(V·s)) |
|---|---|---|---|

| A-2 | 50nm, direct growth with substrate at 375°C | 2.98±0.06 | 687.62±14.59 |
|---|---|---|---|
| C-5 | 50nm: 10nm seed at 350°C, rest at 425°C | 2.08±0.04 | 259.61±3.14 |
| D-4 | 50nm: pre-dose at 450°C, growth at 375°C | A: 3.07±0.02<br>B: 2.99±0.01 | A: 628.77±2.59<br>B: 642.60±1.45 |
| D-5 | 50nm: pre-dose at 440°C, growth at 375°C | A: 3.10±0.03<br>B: 3.09±0.01 | A: 623.71±2.73<br>B: 620.05±1.08 |
| D-6 | 50nm: pre-dose at 425°C, growth at 375°C | 2.99±0.04 | 617.68±7.55 |
| E-2 | 7nm Se sacrificial + 2nm $Bi_2Se_3$ at 75°C, 48nm $Bi_2Se_3$ at 350°C. A and B are both 1cm$^2$ square chips cleaved from the quarter 2-inch wafer | A: 3.98±0.01<br>B: 4.04±0.01 | A: 271.61±0.21<br>B: 269.11±0.30 |
| F-5 | 50nm $Bi_2Se_3$ growth on 50nm BIS at 300°C, BIS following F-4.<br><br>F-4: 10nm BIS seed and 40nm BIS co-deposited at 300°C. Note F-1 to F-4 are all insulating at room temperature. | A: 1.28±0.00<br>B: 1.29±0.00 | A: 930.08±1.33<br>B: 883.50±1.71 |

## 4. Discussion

Overall, we observe very different behaviors for films grown on sapphire and GaAs, likely caused by the absence or presence of dangling bonds. This highlights the importance of the film/substrate interaction even in vdW epitaxy. $Bi_2Se_3$ epitaxy on a substrate with dangling bonds (GaAs in this study) shows far more complicated mechanisms than the classical vdW epitaxy observed in layered chalcogenide materials grown on passivated substrates. For growth on sapphire substrates, the direct growth method results in films with moderate carrier concentrations and mobilities, with relatively small variations in these parameters for different pre-growth treatments or growth strategies. The two-step growth method can reduce the carrier density and improve the mobility at the same time. This can be attributed to the ability to grow the majority of the film at a higher growth temperature, improving adatom surface mobility and resulting in larger domains and a smoother film surface. Unfortunately, the highest deposition temperature for the second step $Bi_2Se_3$ growth is limited by the thermal decomposition point. We tried increasing the second step growth temperature to 450°C but failed, because even though we continuously supplied bismuth and selenium to the $Bi_2Se_3$ seed layer, the materials did not stick. This is confirmed by the RHEED pattern, which shows the sapphire RHEED pattern increasing in intensity over time. This means the material re-evaporation rate is higher than the deposition rate at this temperature. It is possible that higher fluxes could overcome the re-evaporation. The two-step growth method is relatively simple, less time-consuming, and requires minimal pre growth treatment. It is likely a good choice for $Bi_2Se_3$ growth on other passivated substrates.

Things are quite different for $Bi_2Se_3$ growth on GaAs: due to the surface dangling bonds and surface structure anisotropy, the $Bi_2Se_3$ thin film epitaxy is very dependent on the substrate conditions. At the early stages of growth, the $Bi_2Se_3$ (0001) orientation and ($10\bar{1}5$) orientation are usually both present. The ($10\bar{1}5$) orientation grows faster at higher substrate temperatures, and its existence can hamper coalescence of (0001)-oriented domains[64] causing holes in the film. In these films, the spiral structures are prominent and have a high density. These is probably caused by the domain height differences when the in-plane (0001) covers the out-of-plane ($10\bar{1}5$) orientations. When we use the selenium sacrificial layer strategy, the triangular domains show very little spiral growth likely because the growth of the ($10\bar{1}5$) orientation is suppressed. Even though we deoxidized the GaAs surface under a selenium overpressure, the configuration of the surface dangling bonds at high temperatures is unclear. More *in-situ* surface elemental analysis is needed to draw a solid conclusion. Our results on GaAs have shown that even with very different lattice structures, high-quality heteroepitaxy is possible in vdW epitaxy.

Although using a low-temperature elemental selenium layer to passivate the GaAs substrate is somewhat successful, some optimization may still be possible. If we replace the selenium sacrificial layer with a BIS buffer layer, the $Bi_2Se_3$ interaction with the substrate is further reduced, leading to a less defective bottom layer. However, the introduction of a BIS buffer layer may hamper the investigation of the intrinsic properties of $Bi_2Se_3$ for some experimental techniques. It is clear from this study that passivating dangling bonds is of utmost importance for the growth of vdW materials on reactive substrates. The film-substrate interaction is, unsurprisingly, much stronger for vdW materials deposited on reactive substrates like GaAs than for inert substrates like sapphire.

We can also compare the electronic properties results we obtained using these methods to existing results in literature. In general, a 50nm undoped $Bi_2Se_3$ film on sapphire grown by MBE normally shows a sheet carrier concentration about $3\times10^{13}cm^{-2}$ and mobility about $6\times10^2$ $cm^2/(V\cdot s)$ at room temperature. We see similar results when we directly deposit $Bi_2Se_3$ on sapphire. An increase in mobility to 925 $cm^2/(V\cdot s)$ has been shown when the temperature drops to 40K, which is primarily attributed to phonon freeze-out[12]. In a twin-suppressed $Bi_2Se_3$ film, the mobility can reach up to 1593$cm^2/(V\cdot s)$ at 10K[36]. Our results show that by applying the two-step strategy and promoting a larger domain formation in $Bi_2Se_3$ thin film, we can obtain mobilites exceeding 850$cm^2/(V\cdot s)$ at room temperature in a 50nm film. So far, the systems that show higher $Bi_2Se_3$ film mobilities are grown on graphene (up to 3400 $cm^2/(V\cdot s)$ for 200nm film at room temperature[75]), on $MoS_2$ (up to 6000 $cm^2/(V·s)$ for 50nm film with selenium capping at 5K[68]), and on $In_2Se_3$-$Bi_{1-x}In_xSe_3$ bilayer buffer on sapphire (up to 16000 $cm^2/(V·s)$ for 50nm film with selenium capping at 1.5K)[57]. It is worth noting that due to the aging effect in ambient conditions, a capping layer may be helpful for preserving the low carrier density and high mobility of the surface state electrons[76]. In all cases, the highest mobilities and lowest carrier densities have been obtained for films grown on passivated substrates or other vdW layers, rather than on reactive substrates like GaAs. While this is useful for many applications, understanding how to grow high-quality films on semiconductor substrates is needed for integration of these materials with existing optoelectronic device structures. Our results have shown that by promoting the larger domain formation in the film, the carriers experience less domain boundary scattering thus their mobility can be greatly improved.

To further improve the vdW material thin film quality, more efforts are needed around: suppressing the twin domain formation and spiral growth; optimizing the interface between the substrate and film; and promoting a layer-by-layer growth mode rather than an island growth mode. It is possible that substrates with step heights comparable to the thickness of a single layer would reduce the spiral formation. The substrate step edges typically originate from the substrate miscut. After high-temperature treatments, the substrate step heights are usually close to the lattice spacing. However, $Bi_2Se_3$ has a quintuple layer thickness of about 1nm, making it difficult to find a material with a similar lattice out-of-plane lattice constant.

## 5. Conclusion

In this paper, we have shown results of $Bi_2Se_3$ thin films growth via MBE using a variety of growth strategies on GaAs and sapphire. Because of the difference in surface passivation, we observe very different growth dynamics for $Bi_2Se_3$ thin films. Growth on self-passivated sapphire enables a wide substrate temperature growth window, and the film quality is relatively insensitive to the substrate pre-growth treatment and extra annealing time. A two-step growth strategy in which the majority of the $Bi_2Se_3$ layer is deposited near the thermal decomposition point was found to generate films with the best electronic performance: a 20% decrease in carrier density and 60% increase in mobility at room temperature compared to the direct growth method. This is primarily caused by the increased adatom mobility at the high substrate temperature. In this case, the low temperature deposition acts as a wetting layer to enable the high-temperature deposition.

Growth on GaAs substrates shows a significant dependence on the substrate temperature and pre-growth passivation status of the epi-ready surface. The un-passivated surface dangling bonds results in growth of both the (0001) and ($10\bar{1}5$) orientations. For specific growth conditions, the (0001) orientation wins out and buries the other orientation. The anisotropy of the 2x4 reconstructed GaAs surface leads to anisotropic bismuth adatom mobility, which eventually leads to favored in-plane epitaxy along the GaAs [011] axis. The direct growth method generates films with carrier concentrations and mobility similar to films on sapphire, while adding a BIS buffer layer cuts the carrier density in half and increases the mobility by 60%. This is a significant result, since growth of vdW films on technologically-important semiconductor substrates is crucial for future integration of these materials with existing optoelectronic devices. Though there are undoubtedly additional techniques that can be used to improve film quality, these results can be used as a starting point to determine the best techniques for growth of other vdW materials on self-passivated or unpassivated substrates.

**Author Information**

Corresponding Author: *E-mail: slaw@udel.edu

**Acknowledgement**

This project is supported by the U.S. Department of Energy, Office of Science, Office of Basic Energy Sciences, under Award Number DE-SC0017801.

# Optimization of the growth of the van der Waals materials Bi$_2$Se$_3$ and (Bi$_{0.5}$In$_{0.5}$)$_2$Se$_3$ by molecular beam epitaxy

*Zhengtianye Wang, Stephanie Law\**

Material Science and Engineering Department, 201 DuPont Hall, University of Delaware, Newark, Delaware, 19716, U.S.A.

Synopsis: van der Waals materials can be grown at a wafer scale by techniques like molecular beam epitaxy. We compare a variety of growth techniques on an inert substrate (sapphire) and a reactive substrate (GaAs) to understand how the substrate choice impacts film quality. The conclusions can be applied to growth of other van der Waals materials on other substrates.

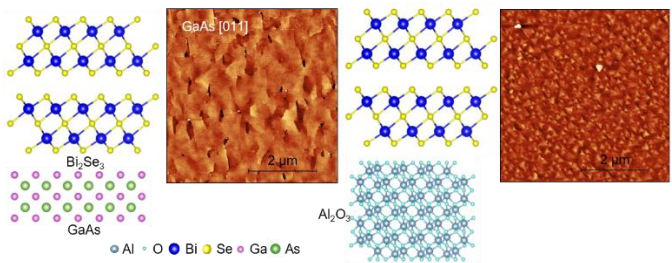